\renewcommand\a{\alpha}
\renewcommand\b{\beta}
\newcommand\g{\gamma}
\renewcommand\d{\delta}
\newcommand\e{\epsilon}

\newcommand\s{\sigma}

\newcommand\f{\phi}

\renewcommand\o{\omega}

\newcommand{\lan}{\langle}
\newcommand{\ran}{\rangle}

\newcommand{\non}{\nonumber\\}

\newcommand{\diracslash}[1]{#1\llap{/\kern2pt}}

\newcommand{\be}{\begin{equation}}
\newcommand{\ee}{\end{equation}}
\newcommand{\bea}{\begin{eqnarray}}
\newcommand{\eea}{\end{eqnarray}}
\newcommand{\ba}[1]{\begin{array}{#1}}
\newcommand{\ea}{\end{array}}
\newcommand{\bep}{\begin{pmatrix}}
\newcommand{\eep}{\end{pmatrix}}

\newcommand{\bt}{\begin{tabular}}
\newcommand{\et}{\end{tabular}}

\newcommand{\beas}{\begin{eqnarray*}}
\newcommand{\eeas}{\end{eqnarray*}}

\documentclass[preprint,prd,aps,floats,nofootinbib,floatfix]{revtex4}
\usepackage{graphicx}
\usepackage[intlimits]{amsmath}
\usepackage[utf8]{inputenc}
\usepackage[T1]{fontenc}
\usepackage{lineno}
\usepackage{textcomp}
\usepackage{subfig}
\usepackage{multirow}
\usepackage{xcolor} 
\usepackage{comment}
\usepackage[a4paper, total={160mm, 240mm}, top={0.5in}]{geometry}
\usepackage{tabularray}
\usepackage{ulem} 
\begin{document}
\title{Meson-nucleus bound states in quark meson coupling model}
\author{Arpita Mondal}
\email{arpita.mondal@physics.iitd.ac.in, arpita.mondal1996@gmail.com
}
\affiliation{Department of Physics, Indian Institute of Technology, Delhi, Hauz Khas, New Delhi -- 110 016, India}

\author{Amruta Mishra}
\email{amruta@physics.iitd.ac.in}
\affiliation{Department of Physics, Indian Institute of Technology, Delhi, Hauz Khas, New Delhi -- 110 016, India} 
\begin{abstract}
The formations of the $K(\bar{K})$, $D(\bar{D})$, and $B(\bar{B})$ meson-nucleus bound states in ${\rm{^{16}O}}$, ${\rm{^{40}Ca}}$, ${\rm{^{90}Zr}}$, ${\rm{^{197}Au}}$, and ${\rm{^{208}Pb}}$ nucleus are investigated using the quark meson coupling model. The model relies on a mean field description of non-overlapping nucleon bags bound by the self-consistent interactions of scalar ($\s$, $\d$) and vector ($\o$, $\rho$) mesons with the (anti)quarks inside the bags, which is further extended to explore the properties of nuclei. We estimate the meson-nucleus bound state energies by solving the Klein-Gordon equations with the real potentials calculated self-consistently within the model, using a coordinate space approach. The calculations are carried out for different nuclear interactions. The effects of Coulomb interaction are considered in the present study for the charged mesons.
Our study indicates the formation of rather deeply bound $B$-mesic states at the very central region of the nuclei, compared to the $D$ and $K$ mesons, offering a more promising probe to explore subtle nuclear medium effects. The investigations of such bound states are of particular interest for the upcoming $\rm{\bar{P}ANDA}$ at FAIR, J-PARC-E29, and JLab experiments.
\end{abstract}
\maketitle

\def\bfm#1{\mbox{\boldmath $#1$}}
\def\bfs#1{\mbox{\bf #1}}
\section{Introduction}
\label{1}
Study of the nuclear many body problem offers a potential direction to understand various aspects of strong interaction physics, including the in-medium properties of hadrons and their interactions with the nuclear matter, and the finite nuclei properties.  
Both experimental \cite{ceres, helios, kek, leps, clas, fopi, kaos1, kaos2, kaos3, kaos4, na50, phenix, e705, alice} and theoretical \cite{Cass, Mishra, hayashigaki,lee,tolos, sarmistha,li97,ko01,J00,fuchs01} studies have shown that under the extreme conditions of nuclear matter, created during the heavy ion collision (HIC) experiments, hadrons undergo in-medium mass drops, attributed to the effective attractive potential experienced by the hadrons within the nuclear medium.
It is suggested that the slow production of these hadrons inside a nucleus, combined with a sufficiently attractive potential, may lead to their absorption and the formation of bound states with the nucleus.
Widely, such states, particularly one with the mesons, are known as “mesic-nuclei”, representing a nuclear system where a meson is bound with a nuclear orbit solely through the strong interaction, without considering the electromagnetic Coulomb effects. The bound state system formed by the combined effects of nuclear and Coulomb interactions is referred to as a “mesic atom”. Notably, the existence of deeply bound pionic states was first discussed by Friedman and Soff \cite{fri}. Following the prediction of ${\pi^-}- {\rm{^{208}Pb}}$ bound states \cite{fri,toki88,toki89,toki91}, the GSI-S160 Collaboration \cite{yama} reported the first observation of the deeply bound pionic states in 1996. The states were revealed as the resonance states in the continuum just below the $\pi^-$ emission threshold in the ${\rm{^{208}Pb}}(d,\, {\rm{^3He}})$ reaction. A detailed discussion of their implications in terms of the pion optical potential and the effective pion mass in the nuclear medium is reported in Refs. \cite{lu,yama98,gigg00,ita00}.
On the other hand, in 1997, Hayano {\textit{et al.}} proposed using the ($d,\, {\rm{^{3}He}}$)
reaction to produce different mesons with nearly
zero recoil \cite{hayano}. If the meson feels a sufficiently large attractive force inside a nucleus, the meson is expected to form meson-nucleus bound states. 
Immense lattice quantum chromodynamics (QCD) calculations \cite{lattice06,lattice10,lattice19} on these states have opened a new type of nuclear spectroscopy: hadron–nucleus bound-state spectroscopy.

The study of mesic-nuclei bound states within the relativistic mean field models provides valuable insights into the meson-nucleon interactions.
Theoretically the formation of such hadronic nucleus states, specifically mesic-nucleus states \cite{lu,pion2,cobos_phi,tsushima_omega_eta,saito_omega,cobos_eta,D1,laura1,brodsky_eta,brodsky_J,QMC_J,bottom} are investigated and their binding energies are estimated through different approaches. The present work focuses on the possibility of obtaining $K(\equiv K^+, K^0),\; \bar{K}(\equiv K^-, \bar{K}^0),\; D(\equiv D^+, D^0),\; \bar{D} (D^-, \bar{D}^0)\; {\rm {and}}\; B(\equiv B^+, B^0 ),\; \bar{B} (\equiv B^-, \bar{B}^0)$ bound states with ${\rm{^{16}O}}$, ${\rm{^{40}Ca}}$, ${\rm{^{90}Zr}}$, ${\rm{^{197}Au}}$, and ${\rm{^{208}Pb}}$ nuclei within the quark meson coupling (QMC) model. Unlike the other hadrodynamics models, here hadrons are no longer treated as point particles. The light quarks and antiquarks inside the non-overlapping nucleons are directly interacting with the mean scalar and vector meson fields. In 1988, Guichon \cite{guichon} proposed the QMC model, inspired by the importance of the internal structure of hadrons on their in-medium properties, as established by the deep inelastic scattering (DIS) experiments and European Muon Collaboration (named the EMC effect) \cite{emc} at Jefferson Laboratory (JLab). 
Within the QMC model, the effects of the nuclear medium as well as the properties of finite nuclei are studied, considering the internal structure response of the hadrons, which offers a good alignment with the experimental findings \cite{krein1,saito2}. The in-medium properties of the mesons constructed only of the light quarks have been discussed thoroughly in Refs. \cite{Mishra, lee, krein1, saito2}. The obtained in-medium masses for the mesons are attributed to the nature of the scalar potentials that the mesons are experiencing within the nuclear matter, which indicates the possible formation of meson-nucleus bound states, motivating further investigation into the properties of these mesons within finite nuclei. Furthermore, in Ref. \cite{patra_delta}, the authors indicated substantial contribution from the inclusion of the scalar-isovector interaction both in finite and infinite nuclear matter, where a possibility of modification of the binding energy, charge radius, and flipping of the orbits in asymmetric finite nuclei has been reported. In Ref. \cite{Saito1994}, the effects of isospin symmetry breaking in the nuclei properties have been investigated by incorporating the scalar-isovector ($\d$) meson within the QMC model to address the Nolen-Schiffer anomaly. The inclusion of the $\d$ (scalar isovector) meson induces the isospin symmetry breaking of the masses of light quark ($u,\;d$) and antiquark ($\bar{d},\;\bar{u}$) doublets, causing the mass splittings within the isodoublets of $K$, $\bar{K}$, $D$, $\bar{D}$, $B$ and $\bar{B}$ mesons in asymmetric medium. The impacts of the scalar-isovector interaction on the properties of pseudoscalar mesons have been studied in Refs. \cite{Saito1994,niu,santos,liu,me} and also reported as a significant contribution to the in-medium properties.
Therefore, our prime interest of this work is to investigate the role of scalar-isovector meson field ${\d}$ along with the isoscalars ($\s$, $\o$) and isovector (${\rho}$) fields to study the probable formation of the bound states with the symmetric (${\rm{^{16}O}}$, ${\rm{^{40}Ca}}$) as well as asymmetric (${\rm{^{90}Zr}}$, ${\rm{^{197}Au}}$, ${\rm{^{208}Pb}}$) nuclei. Contrary to the symmetric nuclear matter scenario, here, the Coulomb interaction induces the asymmetry between the proton and neutron (scalar) densities
for symmetric finite nuclei, which serves a very small but nonzero source for the isovector fields ($\d$ and $\rho$). The study of open strange and open heavy flavor mesons, the systems with single light (anti)quarks, would also provide significant insights into the in-medium dynamics of such systems, offering important information concerning the (partial)restoration of chiral symmetry and the vacuum structure of QCD.  

In Ref. \cite{lee}, Lee and Ko mentioned that the decay width of heavy flavor mesons increases when they are experimentally produced inside a heavy nucleus, resulting in a reduced lifetime in the nuclear matter environment. Therefore, an appreciable fraction of produced meson states are expected to decay inside the nucleus, causing an observable shift in the peak positions in the final particle yields. A probable mechanism of such a physical process of heavy meson-nuclear bound states has been illustrated in Ref. \cite{laura1}. However, it is difficult to observe such bound states experimentally due to their large width compared to the level separation. The J-PARC-E29 \cite{e29i} collaboration has proposed studying in-medium mass modification through the possible formation of mesic-nucleus bound states via antiproton annihilation. Similarly, the Jefferson Lab \cite{jlab1} aims to produce low-momentum mesons within a nucleus. The experimental program $\rm{\bar{P}ANDA}$ at the future Facility for Antiproton and Ion Research (FAIR) \cite{fair1} in Germany plans to investigate bound states up to the charm quarks range. Depending upon the detector capabilities and the energy range of the storage ring, $\rm{\bar{P}ANDA}$ could potentially include heavier quark sectors in its program \cite{hesr}.
In this way, one may indeed anticipate the formation of a mesic-nucleus bound state, where a slowly moving meson is trapped inside the nucleus, where the maximum nuclear matter effect or a minimal variation of the nuclear matter properties can be probed.

The paper is organized as follows:  To begin with, Sec. \ref{2A} provides an overview of the QMC model for finite nuclei, where it describes the interactions within the nucleus. In Sec. \ref{2B} we discuss the mesic-nucleus potentials and the Klein-Gordon equation while the $K$, $\bar{K}$, $D$, $\bar{D}$, $B$ and $\bar{B}$ mesons are captured inside the nucleus. Then we present results and discussions in Sec. \ref{result} for the mesic-nuclei bound states, starting with the discussions on the parameters in Sec. \ref{4A}. In Sec. \ref{4B}, the mean field potentials within the nuclei are studied, while Sec. \ref{4C} briefs the potentials that the mesons are experiencing within the nucleus. Finally, Sec. \ref{4D} analyzes binding energies and discusses various mesic nuclei states. Section \ref{summary}, concludes with a summary of our findings in the current study.

\section{QUARK MESON COUPLING MODEL}
\subsection{Study of finite nuclei}
\label{2A}
A qualitative review of the QMC model can be found in Refs. \cite{krein1,saito2}. The quarks and antiquarks inside the hadrons, which are regarded as static spherical MIT bags, are described as Dirac particles, interact directly with the scalar-isoscalar $\s$ ($J^P=0^+,T=0$), vector-isoscalar $\o$ ($1^-,0$), scalar-isovector $\d$ ($0^+,1$), and vector-isovector $\rho$ ($1^-,1$) meson fields to give rise the nuclear bulk properties. Both of the isovectors $\d$ and $\rho$ mesons simulate the isospin asymmetry of the nuclear matter within the model. More specifically, the $\d$ meson arises due to the mass asymmetry of the nucleons, while the $\rho$ meson addresses the asymmetry in number densities. Using the Born-Oppenheimer approximation, the QMC model can be extended to describe the phenomena of finite nuclei if the quarks inside the nucleons are highly relativistic and the meson fields do not vary rapidly across the interior of the nucleon in a nucleus \cite{guichon1996}. For a complete understanding of nuclear properties, the Coulomb interaction between protons is required to be introduced in the study by the electromagnetic four-potential $A_{\mu}$. The corresponding Lagrangian density is given by
\begin{align}
    \label{A1}
    \mathcal{L}  &=  \bar{\psi}_N\Big[i\gamma.\partial - \Big(m_N - \tilde{g}_\s({\s}) {\s} - \tilde{g}_\d({\d})\frac{{\tau}^a}{2}{\d}^a\Big) - \gamma^\mu\Big(g_\o {\o}_\mu + g_\rho\frac{{\tau}^a}{2}{{\rho}}_{\mu} ^a + \frac{e}{2}(1+\tau^a)A_{\mu} \Big)\Big]\psi_N\nonumber\\ &+\frac{1}{2} ( \partial_\mu {\sigma}\partial^\mu{\sigma} - 
m_\sigma^2{\sigma}^2  ) + \frac{1}{2}(\partial_\mu{{\delta}}^a \partial^\mu {{{\delta}}^a}- m_\delta^2 {{{\delta}}^a}{{{\delta}}^a})- \left[\frac{1}{4}\o_{\mu\nu} \o^{\mu\nu} - \frac{1}{2}m_\omega^2 {\omega}_\mu {\omega}^\mu\right]\nonumber\\ &- (\frac{1}{4}{{\rho}}_{\mu\nu} ^a{{\rho}}^{\mu\nu, a} - \frac{1}{2}m_\rho^2 {{\rho}}_{\mu}^a{{{\rho}}^{\mu,a})} - \frac{1}{4}A_{\mu\nu}A^{\mu\nu},
\end{align}
where,
\begin{align}
\psi_N = \left(\begin{array}{cc}
    \psi_p       \\
    \psi_n       
\end{array}\right),\;\;\;\;\;
m_N = \begin{pmatrix}
     m_p & 0 \\
    0 &  m_n
    \end{pmatrix};
\end{align} 
and $m_\s$, $m_\o$, $m_\d$, and $m_\rho$ are the meson masses and $\tau^a/2$ is the isospin operator for the nucleons with the index $a={1,2,3}$. In the Lagrangian (\ref{A1}), the field strength tensors corresponding to the vector meson fields ($\o$,$\rho$) and the photon fields ($A$) are, respectively,
\begin{align}
\label{tensor}
\omega_{\mu\nu} = \partial_{\mu}{\omega}_{\nu} - \partial_{\nu}{\omega}_{\mu}~, \\
\rho_{\mu\nu} ^a = \partial_{\mu}{{\rho}}_{\nu} ^a - \partial_{\nu}{{\rho}}_{\mu} ^a~,\\
A_{\mu\nu} = \partial_{\mu}{A}_{\nu} - \partial_{\nu}{A}_{\mu}~.
\end{align}
The coupling strengths $\tilde{g}_\s({\s})$ and $\tilde{g}_\d({\d})$ correspond to the nonlinear scalar-meson-nucleon interactions and the vector-meson-nucleon couplings are denoted by $g_\o$ and $g_\rho$. Under the mean field approximation (MFA), the quantum fluctuations of the meson fields are considered to be negligible and the meson fields are treated as classical fields, which amounts to  ${\s} \rightarrow \lan{\s}\ran \equiv \s,\;{{{\d}}^a} \to \lan{{{\d}}^a}\ran \equiv \d^{a3}\d^a,\; {\o}_\mu \rightarrow\lan{\o}_\mu\ran \equiv \d^{\mu 0}\, \o_\mu,\; {{\rho}}_{\mu}^a \to \lan{\rho}_{\mu}^a\ran \equiv \d^{\mu 0} \d^{a 3} \rho_{\mu}^a,\; A_\mu \rightarrow\lan{A}_\mu\ran \equiv \d^{\mu 0}\, A_\mu,$ where only the timelike components of the vector and photon fields will survive due to the time-reversal symmetry of spherical nuclei. As a consequence of the above approximation, the nucleons are moving independently in the mesonic mean field potentials $V_\sigma(r) = g_{\s} (\s(r)) \s(r)$, $V_{\delta} (r) = g_{\delta} (\d(r))\delta^{3}(r)$, $V_{\omega}(r) = g_\omega \omega_0(r)$, $V_{\rho}(r) = g_{\rho}\rho_{0}^3(r)$ and $V_c(r) = eA_0(r)$, satisfying the following Dirac equation, obtained by varying the Lagrangian density (\ref{A1}),
\begin{align} \label{nucleon}
   \Big[i{\g.\partial} - \Big(m_N - V_{\s}(r) - \frac{\tau^3}{2} V_{\d}(r)\Big) - \g^0\Big(V_\o(r) + \frac{\tau^3}{2} V_\rho(r) + \frac{1}{2}(\tau^3+1)V_c(r)\Big)\Big ]\psi_N(x)  = 0. 
\end{align}
Since in Eq. (\ref{nucleon}), the meson fields are classical and the nucleons are operators, the Eq. (\ref{nucleon}) is linear. Hence by expanding in its normal mode solution ($\psi(x)\sim \psi(r) e^{-iEt}$) we obtain the following eigenvalue equation,
\bea
\label{diracnucl}
\{-\a.\nabla + V(r) + \b(m_N+S(r))\} \psi_N(r) = E\psi_N(r)
\eea
where 
\bea
\label{sv}
S(r) &=& V_\s(r) + \frac{\tau^3}{2} V_{\d}(r),\non
V(r) &=& V_\o(r) + \frac{\tau^3}{2} V_\rho(r) + \frac{(1+\tau^3)}{2} V_c(r)
\eea
and the nucleon field operator has the form,
\bea
\hat{\psi}_N(r) = \sum_\a [A_\a\varphi_\a(r) + B^\dagger_\a\zeta_\a(r)],
\eea
which signifies that Eq. (\ref{diracnucl}) has both the positive [$\varphi(r)$] and negative [$\zeta(r)$] energy solutions with $A_\a^\dagger$ and $B_\a^\dagger$ as the creation operators of the nucleon and its anti-particle, respectively. The label $\a$ denotes the single particle orbits \cite{brock,serot}. In this work, we have neglected the Dirac sea contributions since it introduces divergence in theory and which needs to be renormalized in a complicated way \cite{serot}. Therefore, only the following positive energy Dirac spinors are considered in this study under the no-sea approximation,

\begin{align}
\label{diracspinor}
\varphi_{\a}^\kappa ({{r}}) = \begin{pmatrix}
    i\frac{G_\a^\kappa (r)}{r} Y^l_{jm} (\theta,\phi)\\
    -\frac{F_\a^\kappa (r)}{r} Y^{\tilde{l}}_{jm} (\theta,\phi)
\end{pmatrix}  \chi_{t_\a} (t), 
\end{align}
where, $iG_\a(r)/r$ and $-F_\a(r)/r$ are respectively the radial part of the upper (large) and lower (small) components, characterized by the single particle quantum numbers $\a$, orbital momentum quantum numbers $l$, angular momentum quantum numbers $j,m$, and $\kappa$ ($=(-1)^{j+l+1/2}(j+1/2)$) \cite{zhou} and the isospin projection $t_\a = \tau_\a^3/2$ with $\chi_{t_\a}$ being two component Pauli spinor and $Y^{l,\tilde{l}}_{jm}$ are the spin spherical harmonics where $\tilde{l} = l+ (-1)^{j+l-1/2}$. From the definition of the probability of finding a nucleon in space, the normalization condition can be translated as, 
\bea
\label{norm}
\int_0^{\infty} dr (|G_\a(r)|^2 + |F_\a(r)|^2) = 1.
\eea
Now in Spherical coordinates the equations of nucleons can be obtained as
\bea
\label{diraceq}
(\frac{d}{dr} + \frac{\kappa}{r}) G_\a(r) - [\e_\a- V(r)+ m_N - S(r)]F_\a(r) = 0~, \non
(\frac{d}{dr} - \frac{\kappa}{r}) F_\a(r) + [\e_\a - V(r) - m_N + S(r)]G_\a(r) = 0~,
\eea
where $\e_\a$ is the energy (related to the single-particle energy $E_\a$ as $\e_\a = E_\a -m_N $). 
The equations for the boson fields ($M = \s,\d^3,\o_0,\rho_0^3, A^0 $) within static, spherically symmetric nuclei can be obtained as
 \bea
 \label{kg}
 (-\frac{d^2}{dr^2}- \frac{2}{r}\frac{d}{dr} + m_{M}^2) M({r}) &=& \mathcal{J}_{M}({r}),
 \eea
 where $m_{M}$ corresponds to the masses of the respective bosons. However, for the Coulomb field $m_{M}$ values as zero. The source terms of Eq. (\ref{kg}) are,
 \begin{equation}
 \label{source}
 \begin{aligned}
 \mathcal{J}_{M}(r) &= \begin{cases} 
 g_\s \Big[\rho^s_p(r) \left(-\frac{\partial m_p^*(r)}{\partial \s}\right) + \rho^s_n(r) \left(-\frac{\partial m_n^*(r)}{\partial \s}\right)\Big] &\mathrm{for\;\s\;field}\\
 \frac{g_\d}{2} \Big[\rho^s_p(r) \left(-\frac{\partial m_p^*(r)}{\partial \d^3}\right) + \rho^s_n(r) \left(-\frac{\partial m_n^*(r)}{\partial \d^3}\right)\Big] &\mathrm{for\;\d\;field}\\
       g_\o \Big[\rho_p(r)+ \rho_n(r)\Big] &\mathrm{for\;\o\;field}\\
       \frac{g_\rho}{2}\Big[\rho_p(r) - \rho_n(r)\Big] &\mathrm{for\;\rho\;field}\\
       e\rho_p(r) &\mathrm{for\;Coulomb\;field}
       \end{cases}
\end{aligned}
\end{equation}
with the effective nucleon masses
\begin{align}
\label{mass}
m_{N = p,n}^*({\sigma},{\d^3}) = m_N - \tilde{g}_\s({\s}){\s} \mp \frac{1}{2}\tilde{g}_\d({\d}) {{\d}}^3
\end{align} 
and the corresponding proton and neutron densities in the Klein-Gordon equations are defined as
\bea
\label{ps_den}
\rho^s_p(r) 
&=&  \sum_{p}^Z \bar{\varphi}_p(r)\varphi_p(r) \non
&=& \sum_{p}^Z \frac{(2j_p+1)}{4\pi r^2}(|G_p(r)|^2 - |F_p(r)|^2),\\
\label{ns_den}
\rho^s_n(r) 
&=&  \sum_{n}^N \bar{\varphi}_n(r)\varphi_n(r) \non
&=& \sum_{n}^N \frac{(2j_n+1)}{4\pi r^2}(|G_n(r)|^2 - |F_n(r)|^2),\\ \label{p_den}
\rho_p(r) 
 &=&  \sum_{p}^Z {\varphi}^{\dagger}_p(r) \frac{(1+\tau^3_p)}{2}\varphi_p(r) \non
&=& \sum_{p}^Z \frac{(2j_p+1)}{4\pi r^2}(|G_p(r)|^2 + |F_p(r)|^2),\\ \label{n_den}
\rho_n(r) 
 &=&  \sum_{n}^N {\varphi}^{\dagger}_n(r) \frac{(1-\tau^3_n)}{2}\varphi_n(r) \non
&=& \sum_{n}^N \frac{(2j_p+1)}{4\pi r^2}(|G_n(r)|^2 + |F_n(r)|^2),
\eea
where $Z$ and $N$ are the proton number and neutron number of the nucleus, respectively, and $\tau^3_N$ is the isospin operator. The densities are evaluated in the no-sea approximation where the sources of the meson fields are evaluated by considering only the positive energy spinors and therefore, the summation is performed only over occupied orbits in the Fermi sea. 
In Eq. (\ref{kg}), the massive fields are solved by using Green's function method \cite{gambhir,meng}, where
\bea
\label{f}
M(r) = \int_0^\infty r'^{2} dr'G_M(r,r')\mathcal{J}_{M}(r'),
\eea
with
\bea
\label{G}
G_M(r,r') = \frac{1}{2m_M}\frac{1}{r r'}(e^{-m_M|r-r'|}-e^{-m_M|r+r'|}),
\eea
and the massless Coulomb field is solved self-consistently within the QMC model. Collectively, Eqs. (\ref{diraceq}) and (\ref{kg}) are solved by an iterative method, starting with an initial guess of Woods-Saxon potential, where the depth of the potentials are chosen according to the potentials obtained from nuclear matter study within the QMC model. As a consequence, the convergence is achieved within 5-6 iterations, using a computational mesh with a maximum radius of 15 $\rm{fm}$ and a mesh size of 0.01 $\rm{fm}$, combining the numerical technique described in Refs. \cite{brock,saito2}.

Within the nucleus, the potentials and their source densities depend on the radial distance $r$.
At the position $x = (t,\vec{r})$ in the nucleus, the
Dirac equations for the constituent (anti)quarks inside the hadron bags ($h=n,p,...$) are represented as,

\bea
\label{uq}
&\Big[i\g.\partial - \Big(m_{u} - V_{\sigma}^q(r) \mp \frac{1}{2}V_{\delta}^q(r) \Big) \mp \g^0\Big(V_{\omega}^q(r) + \frac{1}{2} V_{\rho}^q(r)\Big)\Big]
\begin{pmatrix}
    \psi_{uh}(x)  \\
    \psi_{\bar{uh}}(x) 
\end{pmatrix} = 0,\\
\label{dq}
&\Big[i\g.\partial - \Big(m_{d} - V_{\sigma}^q(r) \pm \frac{1}{2}V_{\delta}^q(r) \Big) \mp \g^0\Big(V_{\omega}^q(r) - \frac{1}{2} V_{\rho}^q(r)\Big)\Big]
\begin{pmatrix}
    \psi_{dh}(x)  \\
    \psi_{\bar{dh}}(x)  
\end{pmatrix} = 0,
\eea
with $V_\sigma^q(r) = g_{\sigma}^q \sigma(r)$, $V^q_{\delta}(r) = g_{\delta}^q\delta^{3}(r)$, $V_{\omega}^q(r) = g_\omega^q \omega_0(r)$, and $V_{\rho}^q(r) = g_{\rho}^q\rho_{0}^3(r)$ are the mean field potentials,
where the fields $\s(r)$, $\d^3(r)$, $\o_0(r)$, and $\rho_{0}^3(r)$ are obtained in a self-consistent way from the Eq. (\ref{kg}). Here, $m_u(=m_{\bar{u}})$, $m_d(=m_{\bar{d}})$, and $m_Q$ refer to the current quark masses and the expressions for the effective light quarks and antiquarks masses can be defined as
\bea
\label{umass}
\left( \begin{array}{c} m_u^\star(r) \\ m_{\bar{u}}^\star(r) \end{array} \right)
&=& m_{u,\bar{u}}-V_\s^q(r) + \left( \begin{array}{c} - \frac{1}{2} \\ + \frac{1}{2} \end{array} \right) V_\d^q(r),\non
\left( \begin{array}{c} m_d^\star(r) \\ m_{\bar{d}}^\star(r) \end{array} \right)
&=& m_{d,\bar{d}}-V_\s^q(r) + \left( \begin{array}{c} + \frac{1}{2} \\ - \frac{1}{2} \end{array} \right) V_\d^q(r).
\eea
The (anti)quark effective masses are modified by the mean scalar potentials. However, the heavy quark ($Q=s,c,b,\bar{s},\bar{c},\bar{b}$ ) masses are assumed to be the same in the nuclear medium as in the free space ($m^\star_{Q} = m_{Q}$), since, in the QMC model, the mean fields are coupled only to the light quarks and light antiquarks.
In Eqs. (\ref{uq}) and (\ref{dq}), $\psi_{fh}$ represents the static ground state wave function for the (anti)quark fields of flavor $f(=q,Q)$ in the bag, and, is given by
\begin{align}
\label{B25}
\psi_{fh} ({x}) &= N_{fh} e^{- \frac{i \epsilon^\star_{fh} t}{ R^\star_h}}\begin{pmatrix}
    j_0(\frac{x^\star_{fh} r}{R_h^\star}) \\
    i\beta^\star_{fh}\vec{\sigma}.\hat{r} j_1(\frac{x^\star_{fh} r}{R^\star_h})
\end{pmatrix} \frac{\chi_{fh}}{\sqrt{4\pi}}, \non
&= N_{fh} e^{- \frac{i \epsilon^\star_{fh} t}{ R^\star_h}}\psi_{fh}(r)
\end{align}
where $N_{fh}$ is the normalization factor \cite{saito2}, $j_0$ and $j_1$ are the spherical Bessel functions, $R_h^\star$ is the in-medium bag radius, and  $\chi_{fh}$ refers to the quark spinors. The energy eigenvalues (in units of $1/R_h^\star$) of the quarks and antiquarks, which are denoted by $\epsilon_{fh}^\star$ in Eq. (\ref{B25}), are given as,
\begin{eqnarray}
\label{B26}
\left( \begin{array}{c} \e^\star_{uh} \\ \e^\star_{\bar{u}h} \end{array} \right)
&=& \Omega_{qh}^\star \pm R^\star_h \left( 
V_\omega + \frac{1}{2} V_\rho(r) \right),\\ \label{B27}
\left( \begin{array}{c} \e^\star_{dh} \\ \e^\star_{\bar{d}h} \end{array} \right)
&=& \Omega_{qh}^\star \pm R^\star_h \left( 
V_\omega - \frac{1}{2} V_\rho \right),\\ \label{B28}
\e^\star_{Qh} &=& \e_{Qh}= \Omega^\star_{Qh}.
\end{eqnarray}
The other parameters in Eq. (\ref{B25}) are 
\begin{align}
\label{B29}
\beta^\star_{fh} = \sqrt{\frac{\Omega_{fh}^\star - R^\star_h m_{fh}^\star}{\Omega_{fh}^\star + R^\star_h m_{fh}^\star}}\;,\;\;\; {\rm{where,}}\;\;\; 
\Omega_{fh}^\star = \sqrt{x_{fh}^{\star 2} + (R^\star_h m_{fh}^\star)^2}.
\end{align}
The parameters $x_{fh}$'s are the bag eigen frequencies in units of $1/R_h^\star$ \cite{MIT} determined by the linear boundary condition at the bag surface \cite{thomas1982}
\begin{align}
    \label{B31}
    i\g^{\mu} n_{\mu} \psi_{fh} = \psi_{fh}\; \implies\;j_0(x^\star_{fh}) = \beta^\star_{fh} j_1 (x^\star_{fh}).
\end{align}
In the QMC model, the effective mass of the hadrons in the nuclear medium at rest is equal to the energy of the static bag consisting of ground state quarks and antiquarks, which can be obtained as,
\begin{align}
    \label{B32}
     m_h^\star(\s,\d^3) =E_h^{bag}(\s,\d^3)= \frac{\sum_f{n_{fh}\Omega_{fh}^\star}-Z_h}{R^\star_h} + \frac{4}{3} \pi R_h^{\star 3} B,
\end{align}
where the effective hadronic masses implicitly depend on position through the scalar fields. In Eq. (\ref{B32}), the parameter $Z_h$ provides the correction regarding the center of mass motion and the gluon fluctuations and $B$ is the bag pressure which balances the outward pressure exerted by the motion of the quarks and antiquarks inside the bag. As a consequence, there is an equilibrium where the hadron mass
is minimized and stabilized, as follows :
\begin{align}
    \label{B33}
    \frac{d m_h^\star(r)}{d R_h}{\Big{|}}_{R_h^\star} = 0.
\end{align}
Using the above condition the in-medium bag radius, $R_h^\star$, can be determined. It is important to remember that the bag radius is different from the actual hadron radius, which can be computed from the quark wave functions. 

 From Eqs. (\ref{kg}) and (\ref{source}), it is observed that the scalar fields (${\f = \s,\d}$) exhibit self-consistent interactions within the nucleus. The scalar fields satisfy the self-consistent relation in the following way:
 \begin{eqnarray} \label{A12}
 - \frac{\partial  m_h^\star(\f(r)) }{\partial \f} &=& \tilde{g}_\f(\f(r)) = g_\f(\f=0)
 C_h(\f(r)),\\
 \label{A13}
 C_h(\f(r)) &=& \frac{\sum_q n_{qh} S_{qh}(\f(r))}{\sum_q n_{qh} S_{qh}(\f=0)},
 \end{eqnarray} 
 where $n_{qh}$ is the number of light valence (anti)quark constituents and $S_{qh}(\f(r))$ represents the scalar density of the light (anti)quarks ($q = u,\bar{u},d,\bar{d}$) of the respective hadron ($h$).
 At the hadronic level, all the information about the constituent quarks is encrypted within this term. It might be noted that, for $C(\s)=1$, the QMC model reduces to the Walecka model \cite{walecka}, where the nucleons are treated as point particles.
 The meson-nucleon couplings $(g_\s,\,g_\d,\;g_\o,\;g_\rho)$ are defined based on 
 quark-meson coupling strengths
 $(g_\s^q,\;g_\d^q,\;g_\o^q\;g_\rho^q)$ as \cite{guichon1996}:
 \begin{eqnarray} \label{A14}
 g_\f (\f =0) &=& g_\f^q \sum_q n_{qh} S_{qh}(\f=0),\;
 \label{A16}
 g_\o = g_\o^q\sum_q n_{qh},\;g_\rho = g_\rho^q.
 \label{gNs}
 \end{eqnarray}

The scalar density $S(\f(r))$, given in Eq. (\ref{A13}) is expressed in terms of the scalar densities of the light (anti)quarks in the bag, which are given as \cite{niu}
\bea \label{B34}
S_{qh}(\s(r))&=& \mathcal{I}^s_h ,\;\;\;\;S_{qh}(\d^3(r)) = \frac{\tau_q^3}{2}\mathcal{I}^s_h,\\
\mathcal{I}^s_h &=& \int d{r}\bar{\varphi}_{qh}(r) \varphi_{qh}(r) 
\nonumber\\ &=&  \frac{\Omega_{qh}(r)/2 + m_{qh}^*(r) R^*_h(\Omega_{qh}(r)-1)}{\Omega_{qh}(r)(\Omega_{qh}(r)-1) + m_{qh}^*(r) R^*_h /2}.
\eea
The above term carries the internal structure responses of the hadrons and thus plays an important role in this model.
\subsection{Mesic-nuclei potentials}
\label{2B}
In this section, we present the
potentials experienced by the mesons ($P\equiv K,\bar{K}, D,\bar{D}, B,\bar{B}$) while they are captured inside the nucleus while moving in a relatively low momentum. Within the model, the quarks inside the mesons are directly interacting with the mean field potentials inside the nucleus. Consequently, the mesic-nuclei potentials are obtained as a combination of the mean-field potentials $V_{\s,\d,\o,\rho}^q(r)$, calculated self-consistently from Eqs. (\ref{diraceq}) and (\ref{kg}). The effective influence of the scalar and vector potentials on the different mesons while in the nucleus can be written in the following way:
\begin{align} \label{C1}
V_s^P(r)& = m_P^\star(r) - m_P,\\
\label{C2}
V_v^{K^-,B^-}(r) &= - V_\omega^q(r) - \tfrac{1}{2} V_\rho^q(r) - A(r),
&
V_v^{D^-}(r) &= V_\omega^q(r) - \tfrac{1}{2} V_\rho^q(r) - A(r), \\
\label{C3}
V_v^{\bar{K}^0,\bar{B}^0}(r) &= - V_\omega^q(r) + \tfrac{1}{2} V_\rho^q(r),
&
V_v^{\bar{D}^0}(r) &= V_\omega^q(r) + \tfrac{1}{2} V_\rho^q(r),\\
\label{C4}
V_v^{K^0,B^0}(r) &= V_\omega^q(r) - \tfrac{1}{2} V_\rho^q(r),
&
V_v^{D^0}(r) &= - V_\omega^q(r) - \tfrac{1}{2} V_\rho^q(r).
 \end{align}
With a careful inspection of the above potentials for the $P$ meson inside a nucleus, i.e., $V_s^P + V_v^P$, one can draw the naive idea of the existence of the $P$-mesic bound states with the considered nucleus. Since the considered mesons represent heavy-light systems, they can be explained by the relativistic Klein-Gordon equation and the energies of the bound state of those mesons in the nucleus can be obtained by solving the following equation (\ref{final}), using the potentials calculated within QMC model, 
\begin{align}
\label{final}
 \big[\nabla^2 + [E_P-V_v^P(r)]^2 - m_P^{\star 2 }(r)\big] \Phi_P(r) = 0, \;\;\; {\rm{where}}\;\;[\nabla^2 = \frac{d^2}{dr^2} - \frac{l(l+1)}{r^2}],
\end{align}
with $\Phi_P(r)$ is the corresponding state wave function and $E_P$ is the total energy of the meson, where the binding energy for the state is defined as $\epsilon_P = (E_P - m_P)$.
\section{Results and Discussions}
\label{result}
In the present work, we study the possible formation of the bound states of open strange ($ K^+,\;K^-,\;K^0,\;\bar{K}^0$), open charm ($D^+,\;D^-,\;D^0,\;\bar{D}^0$), and open beauty ($ B^+,\;B^-,\;B^0,\;\bar{B}^0$) mesons with ${\rm{^{16}O}}$, ${\rm{^{40}Ca}}$, ${\rm{^{90}Zr}}$, ${\rm{^{197}Au}}$, and ${\rm{^{208}Pb}}$ nucleus within the QMC model. 
Using the obtained meson–nucleus potentials from Eqs. (\ref{C1})-(\ref{C4}) and the in-medium masses from Eq. (\ref{B32}), we solve the Klein-Gordon equation and calculate the binding energies for the mesic–nucleus states. In the following subsections we present the parameters of the model, the behavior of the mean field potentials with their source density distributions within the nuclei, demonstrate the behavior of the meson masses inside the nucleus as well as mesic nucleus potentials, and the ground state eigen-energies associated with the bound states.
\subsection{Parameters of QMC model} \label{4A}
We first state the parameters of the model we have chosen in the present work before discussing the results. The bag parameters $B$ and $Z$ are fitted from the mass and the given bag radius of the proton in the free space, using Eqs. (\ref{B32}) and (\ref{B33}). It might be
noted here that, in the present study, we have considered the small isospin breaking effect between charged and neutral hadrons due to the electromagnetic corrections.  Therefore, by fixing the bag constant, $B = (170.2416\; \rm{MeV})^4$, the values of $R$ and $Z$ are calculated for different mesons with different quark contents and their respective masses (in {$\rm{MeV}$}), i.e., $m_u = 2.16$, $ m_d = 4.67$, $ m_s = 93.4$, $m_c = 1270$, and $m_b = 4180$ \cite{pdg}. The obtained parameters are shown in Table \ref{t1}.
\begin{table}[th]
\centering
\begin{tblr}{colspec={ccccccc}}
\hline
& $M$(MeV) (\textit{I}) &$R$ (fm)& $Z$\\
\hline
$p$ & 938.272&  0.8 (\textit{I}) &  3.2925\\
$n$&  939.565 &  0.8004 & 3.2921\\
\hline
$K^0\; ({\bar{K}^0})$ &  497.611&  0.6419 & 3.0210\\
$K^+\; (K^-$) &  493.677 &  0.6402 &  3.029\\
\hline
$D^0\; ({\bar{D}^0})$ &  1864.84&  0.7494 &  1.2977\\
$D^+\; ( D^-)$ &  1869.66 &  0.7508 &  1.2843\\
\hline
$B^0\; ({\bar{B}^0})$ &  5279.66&  0.8580 &  -1.2161\\
$B^+\; ( B^-)$ &  5279.34 &  0.8579 &  -1.2200\\
\hline
\end{tblr}
\caption{ \raggedright{Representative inputs (\textit{I}) and parameters that are used in the present study.}}
\label{t1}
\end{table}
\begin{table}[th]
\centering
\begin{tabular}{ c   c   c   c   c   c}
\hline
 $g_\s$ & \qquad $g_\o$ & \qquad $g_\d$ & \qquad $g_\rho$& \qquad$m_\s$ (MeV) & \qquad $m_\d$ (MeV)\\
\hline
6.26 &\qquad  8.17 &\qquad 18.63 &\qquad  9.33 &\qquad  418 &\qquad  1000\\
\hline
\end{tabular}
\caption{ \raggedright{Values of nucleon-meson-field coupling constants as well as $m_\s$ and $m_\d$.}}
\label{mod_param}
\end{table}

Now, following Ref.~\cite{saito2}, we fix the parameters $g_\s,\;g_\d,\;g_\o, {\rm{and}}\;g_\rho$ and $e,\; m_\s,\;m_\d,\; m_\o\; {\rm{and}}\; m_\rho$. The meson nucleon coupling strengths $g_\s$, $g_\d$, $g_\o$, and $g_\rho$ are adjusted through the nuclear saturation phenomena and the bulk symmetry energy per baryon with the parameters $m_\s = 550\;\rm{MeV},\;m_\d = 980\;\rm{MeV},\;m_\o = 783 \;\rm{MeV}$, $m_\rho = 770 \;\rm{MeV}$. However, the $m_\s$ affects the finite nuclei properties significantly.
The $m_\s$ is fitted through the root mean square (rms) charge radius of ${\rm{^{40}Ca}}$ \cite{caexp,saito1} and $g_\s$ is calculated by keeping $(g_\s/m_\s)$ fixed. We further assume $(g_\d^{FN}/g_\d^{NM})=(g_\s^{FN}/g_\s^{NM})$, where `FN': finite nucleus and `NM': nuclear matter, in order to obtain $g_\d$ for finite nuclei, from which $m_\d$ is determined so that the ratio $(g_\d/m_\d)$ is maintained same as in nuclear matter. The parameters for nuclear interactions are summarized in Table \ref{mod_param}.
Moreover, the electromagnetic coupling strength is fixed using the experimental value $e^2/4\pi(=1/137.036)$. In the present work, we also examine the effects of a larger vector coupling, where $g_\o^q$ is rescaled by a factor of $(1.4)^2$. This rescaling is based on the empirically obtained $K^+ N$ scattering data \cite{tsushima1}, which is also applicable for the open charm and open beauty mesons \cite{D1}.
\subsection{Mean (scalar and vector) field potentials in nucleus} \label{4B}
\begin{figure}[th]
\centerline{\includegraphics[width=\textwidth]{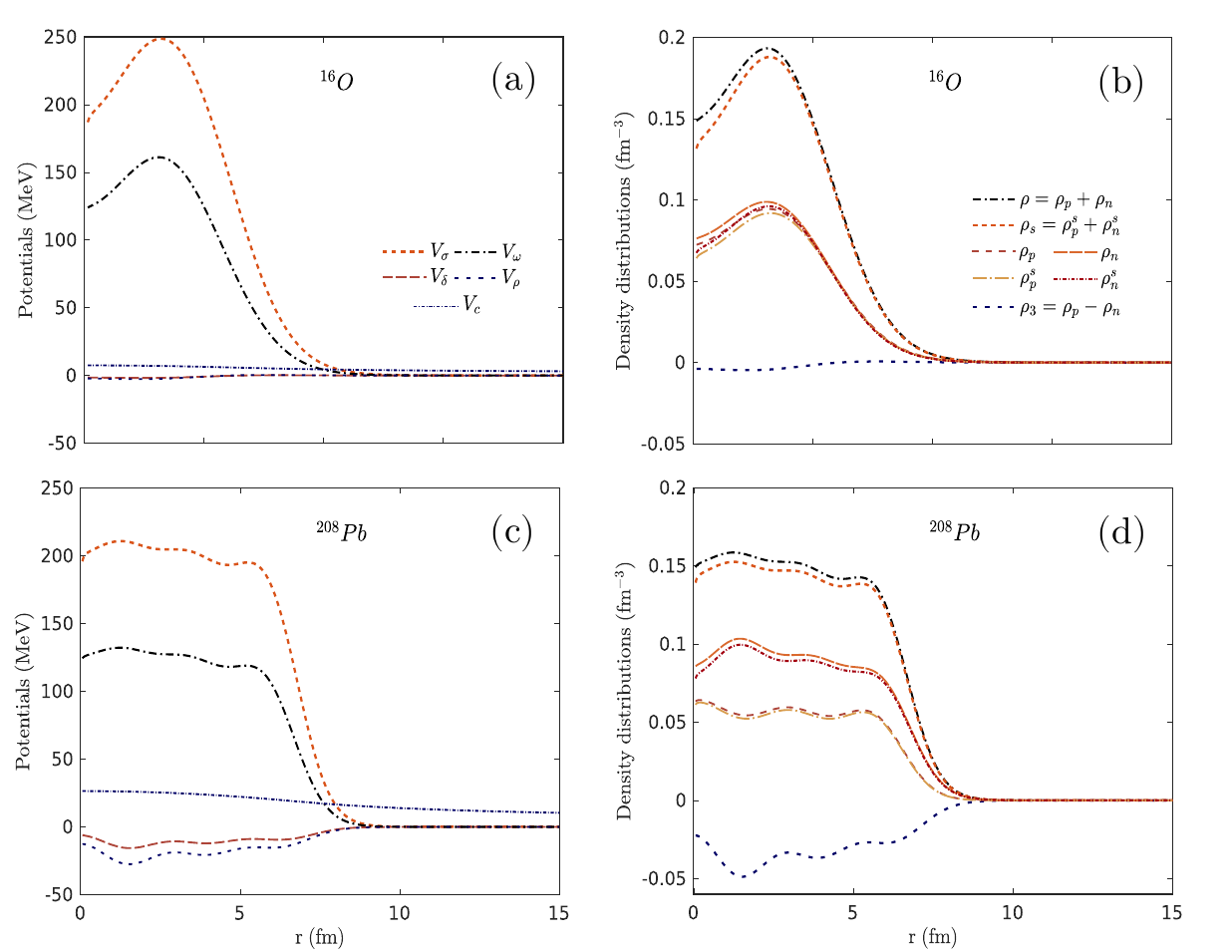}}
\vspace*{8pt}
    \caption{\raggedright{Behaviors of (i) the mean field (scalar-isoscalar, scalar-isovector, vector-isoscalar, and vector-isovector) potentials as well as Coulomb potential, and (ii) the source density distributions within the nucleus. Here we present results for two nuclei:  (a),(b) the symmetric nucleus ${\rm{^{16}O}}$, and (c),(d) the asymmetric nucleus ${\rm{^{208}Pb}}$.}}
    \label{fden}
\end{figure}

We first examine the behavior of the relativistic mean field (scalar and vector) potentials and the Coulomb potentials across various nuclei. Using the parameters discussed in the earlier subsection, the mean field potentials are determined self consistently by solving the coupled differential Eqs. (\ref{kg})-(\ref{n_den}) combining the numerical techniques demonstrated in \cite{brock,serot,gambhir}. The obtained findings are illustrated in Figs. \ref{fden}(a,c), where we present the results for the symmetric nucleus ${\rm{^{16}O}}$ and for the asymmetric nucleus ${\rm{^{208}Pb}}$, defined in terms of their number of constituent nucleons.
The figures indicate that the potentials $V_\s(r)$, $V_\d(r)$, $V_\o(r)$, and $V_\rho(r)$  are largely dictated by their source densities, as shown in Figs. \ref{fden}(b,d).
The total baryon density and as a consequence, the scalar and vector potentials are nearly uniform in the interior of the nucleus and fall off smoothly at the nuclear surface.
At the center of a nucleus, the predictions of finite nuclear study are in firm agreement with the empirically obtained central density or the saturation density of the nuclear matter, $\rho_0\sim0.15\;{\rm{fm}}$, and the associated nuclear matter potentials \cite{saito1, me}. 
The observed difference between the scalar and the vector potentials is manifested by the different signs of the lower components of nucleon spinors in the respective densities, see Eqs. (\ref{source}) and (\ref{ps_den})-(\ref{n_den}). The isovectorial potentials $V_\d(r)$ and $V_\rho(r)$ emerge due to the disparity between proton and neutron distribution, as shown in Figs. \ref{fden}(b,d). In nuclei such as ${\rm{^{16}O}}$ with equal numbers of protons and neutrons, the Coulomb interaction causes the isovectorial interactions to attain small nonzero values. However, for neutron-rich nuclei such as ${\rm{^{208}Pb}}$, $V_\rho$ attains more negative expectation values compared to symmetric nuclei, as illustrated in Fig. \ref{fden}(c). The Coulomb potential $V_c$, depicted in Figs. \ref{fden}(a,c), results solely from the proton distribution within the nucleus. Similar calculations are also performed for the following nuclei, ${\rm{^{40}Ca}}$, ${\rm{^{90}Zr}}$, and ${\rm{^{197}Au}}$.
Notably, we found a good agreement of the obtained density distributions of the nuclei with the experimental results provided by PREX Collaboration \cite{prex,pden}.
\subsection{$K(\bar{K})$, $D(\bar{D})$ and $B(\bar{B})$ mesic-nuclei potentials} \label{4C}

In this subsection, we discuss the mesic-nuclei potentials when $K(\bar{K})$, $D(\bar{D})$, and $B(\bar{B})$ mesons are produced inside a light as well as a heavy nucleus with relatively low momenta. 
\begin{figure}[th]
\centerline{\includegraphics[width=14cm, height = 17cm]{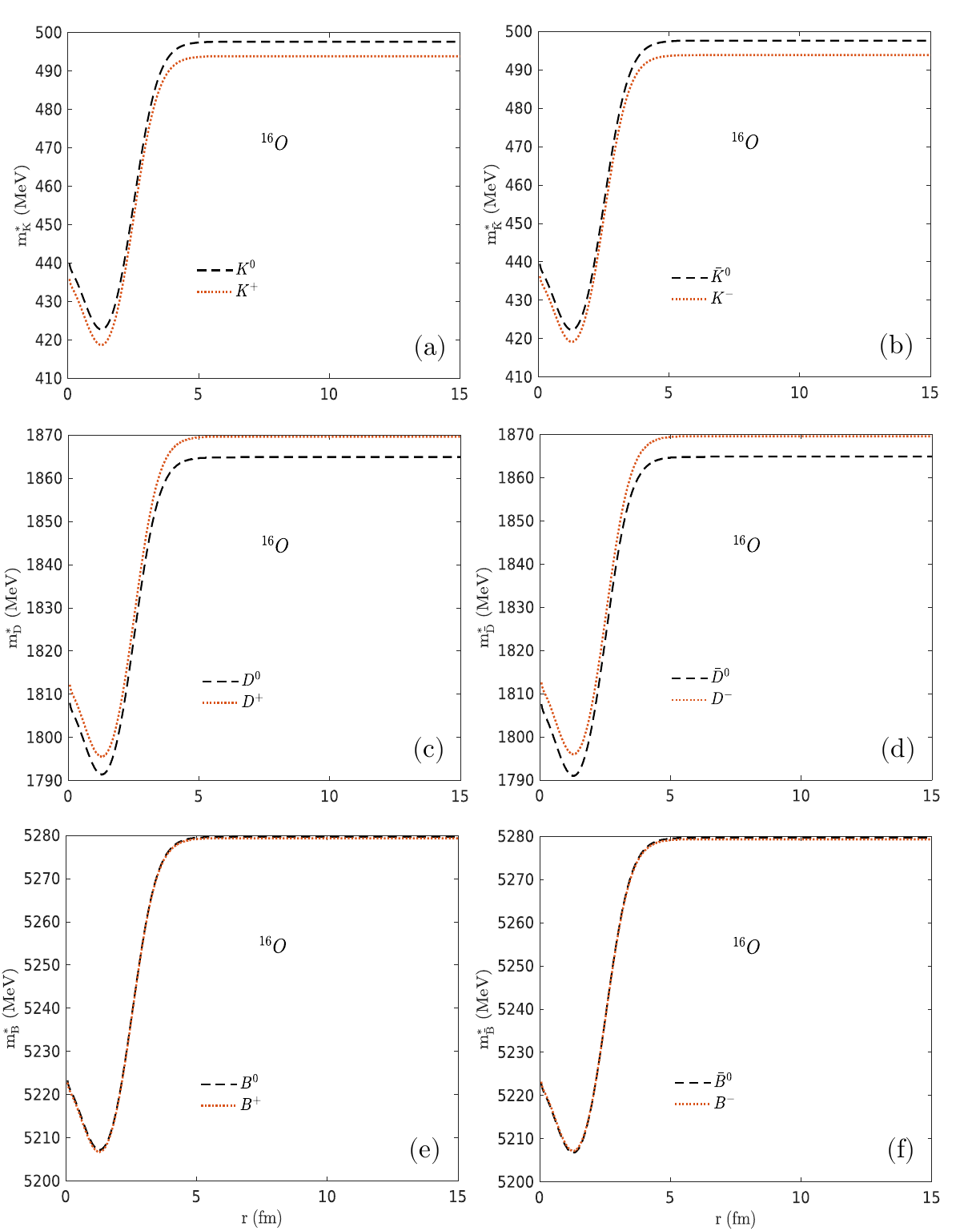}}
\vspace*{8pt}
    \caption{\raggedright{The effective masses of (a) $K$, (b) $\bar{K}$, (c) $D$, (d) $\bar{D}$, (e) $B$, and (f) $\bar{B}$  mesons  while captured inside the ${\rm{^{16}O}}$ nucleus.}}
    \label{mass_o}
\end{figure}
\begin{figure}[th]
\centerline{\includegraphics[width=14cm, height = 17cm]{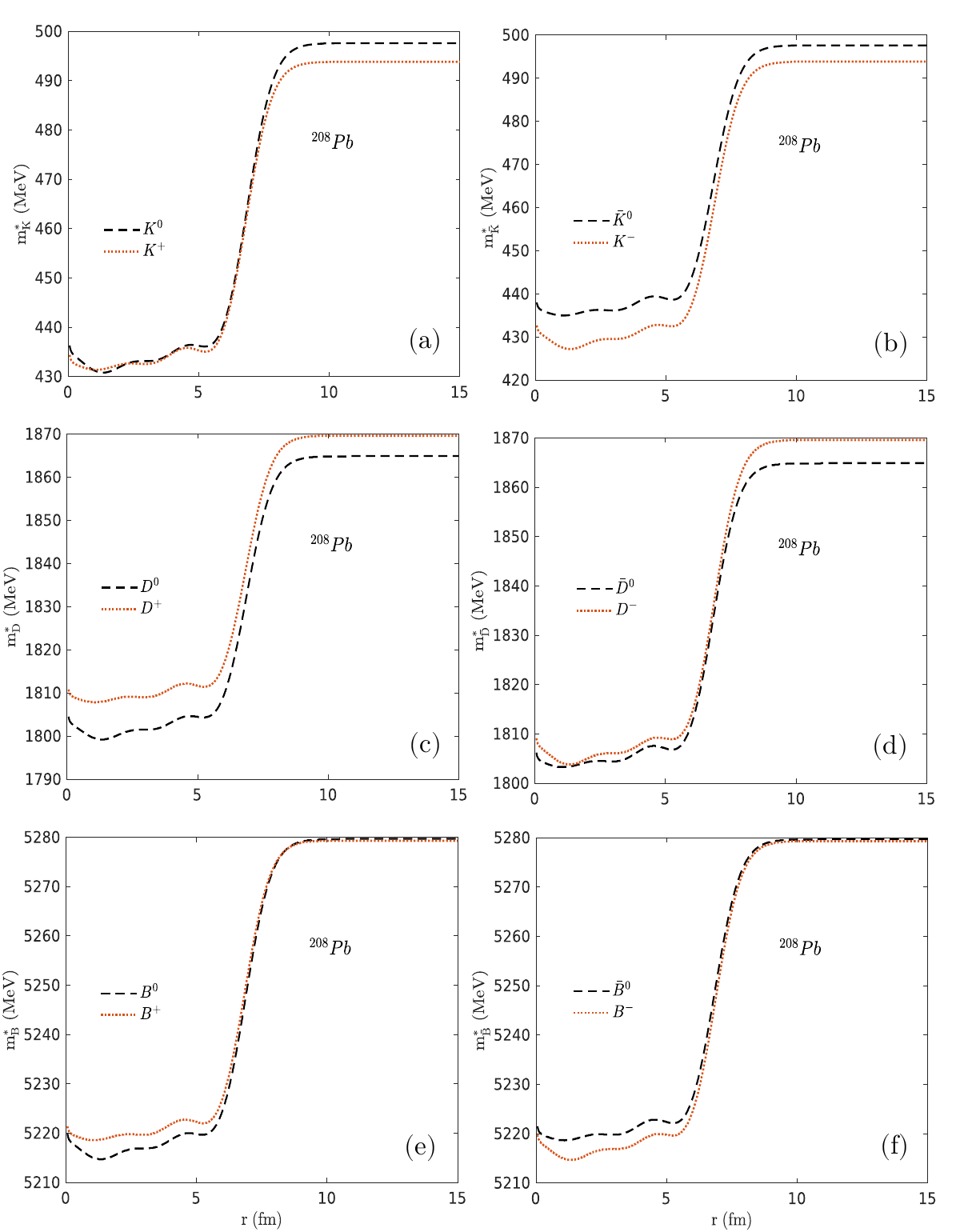}}
\vspace*{8pt}
    \caption{\raggedright{Same as Fig. \ref{mass_o}, for ${\rm{^{208}Pb}}$ nucleus.}}
    \label{mass_pb}
\end{figure}
As stated earlier, within the QMC model, the light quarks and antiquarks constituents of the mesons directly interact with the mean field potentials developed within the nucleus. The interaction of the light quarks and antiquarks with the scalar mean field potential of the nuclei modifies the (anti)quark masses [see Eqs. (\ref{umass})] and most importantly the consideration of the $\d$ field in this study, breaks the mass degeneracy of the isodoublet components of quarks and antiquarks i.e., between ($u,\;d$) as well as ($\bar{d},\;\bar{u}$). Such considerable mass splittings can be understood from Eq. (\ref{umass}), where
the $\d$ meson induces opposite interactions between the (anti)quarks within the isospin doublets and hence leads to a repulsive contribution to the $u(\bar{d})$ mass and attractive contribution for $d(\bar{u})$ mass. As a result, the mass drop of the $u$ quark becomes lesser than that of the $d$ quark, and the mass drop of $\bar{u}$ quark becomes higher than the mass shift of $\bar{d}$ quark, which are usually identical to its antiparticle when the isovectorial interaction is not taken into account.
This further will modify the masses of the open strange and open heavy flavor mesons through the relation in Eq. (\ref{B32}), which are shown in Figs. \ref{mass_o} and \ref{mass_pb}.
One can see from the figures that, the effective masses of the members in the isospin doublets experience splittings in the nuclei due to the consideration of the scalar-isovector interaction, which induces significant splittings for heavier nuclei.

\begin{figure}[th]
\centerline{\includegraphics[width=\textwidth]{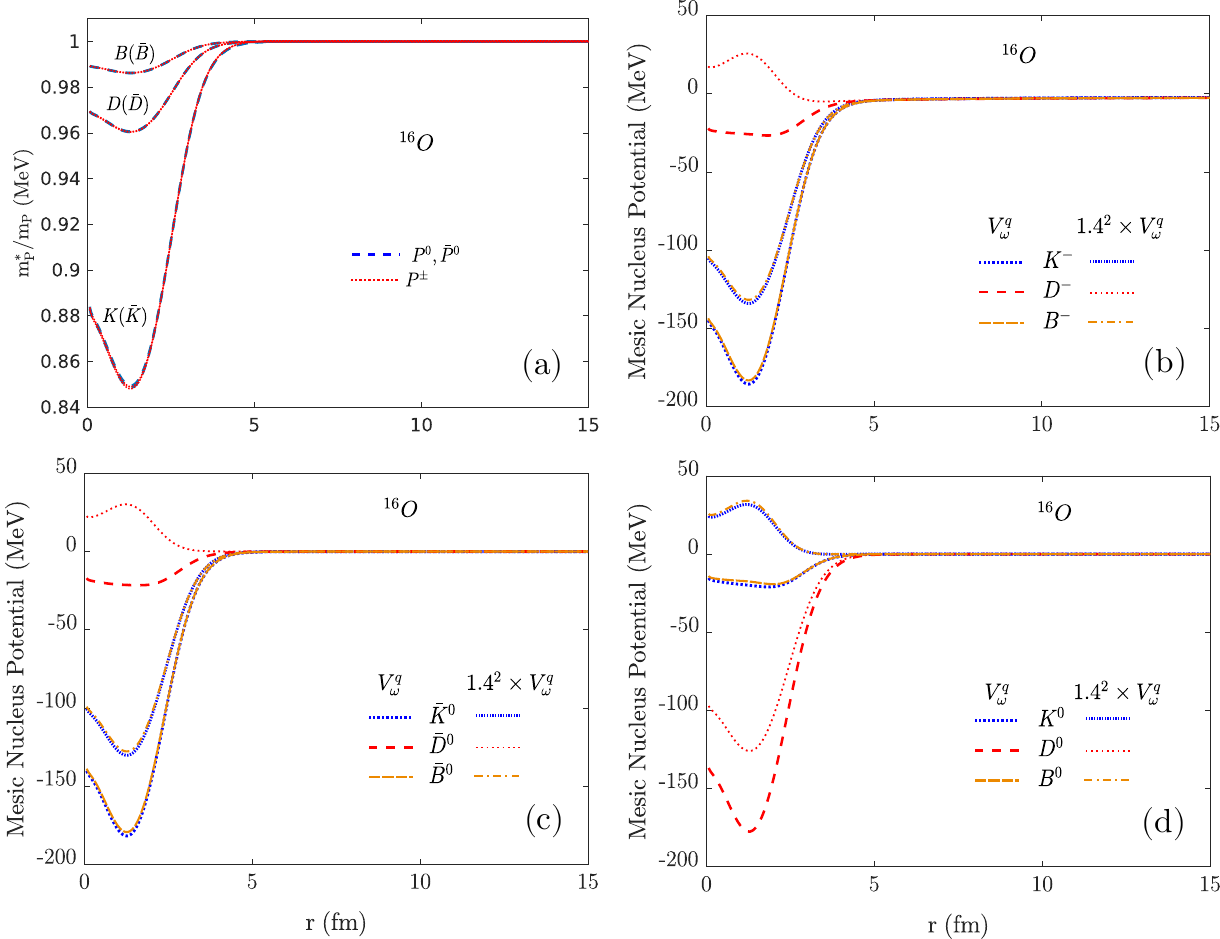}}
\vspace*{8pt}
    \caption{\raggedright{Behavior of (a) the relative masses and (b)-(d) the vectorial interactions of $P^-$ ($K^-$, $D^-$, $B^-$), $\bar{P}^0$ ($\bar{K}^0$, $\bar{D}^0$, $\bar{B}^0$), and $P^0$ ($K^0$, $D^0$, $B^0$) mesons while captured inside the ${\rm{^{16}O}}$ nucleus. Results are presented for both the standard $\omega$ coupling ($V_\omega^q$) and the phenomenologically suggested $\omega$ coupling ($\tilde{V}_\omega^q = 1.4^2 \times V_\omega^q$).}}
    \label{pot_o}
\end{figure}
\begin{figure}[th]
\centerline{\includegraphics[width=\textwidth]{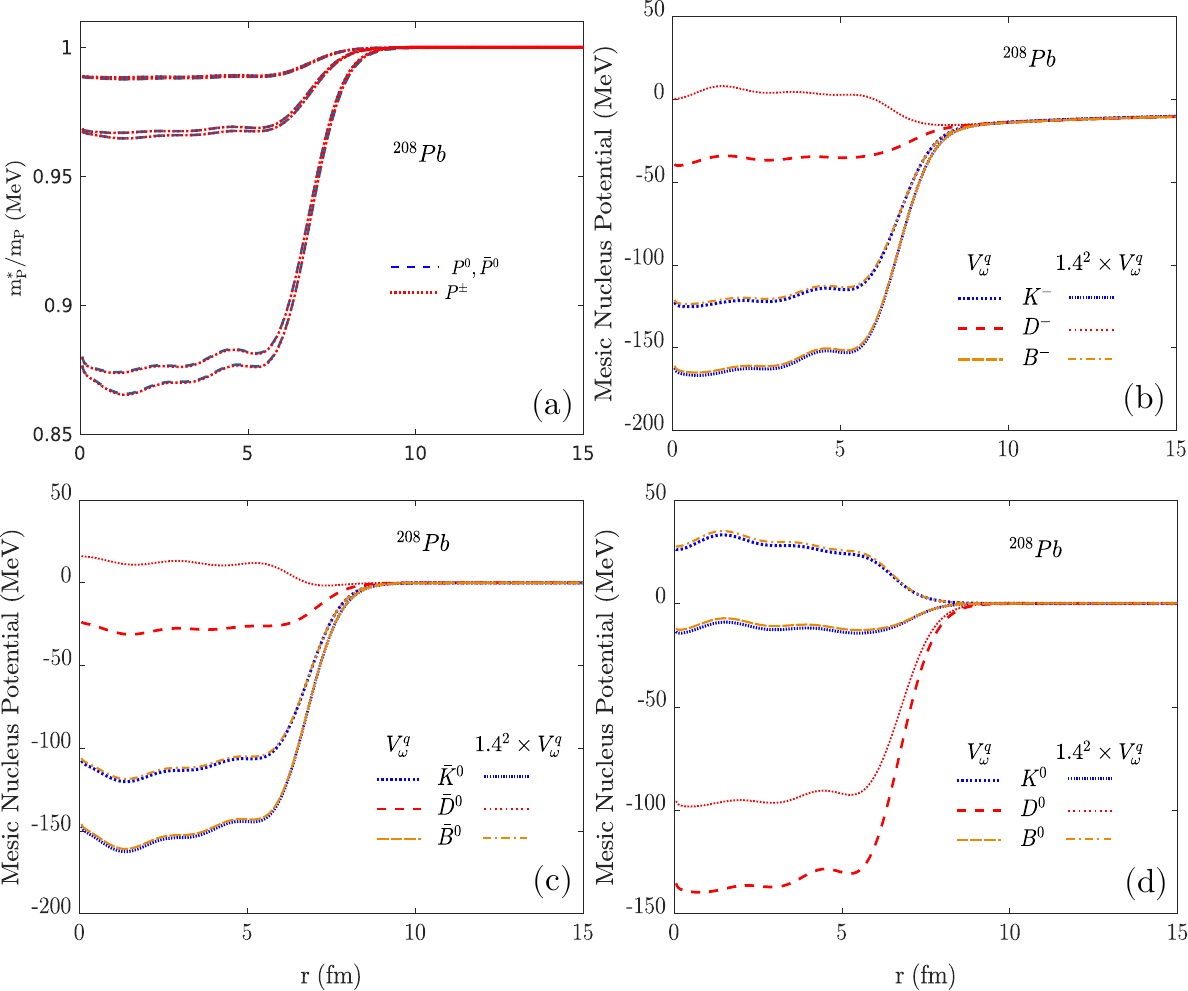}}
\vspace*{8pt}
    \caption{\raggedright{{Same as Fig. \ref{pot_o}, for ${\rm{^{208}Pb}}$.}}}
    \label{pot_pb}
\end{figure}
It is important to mention that as one approaches the center of the nucleus, the average baryon density of the nucleus converges to the nuclear matter saturation density ($\rho_0\sim0.15 \;{\rm{fm^{-3}}}$). Consequently, the results obtained in the interior of the nucleus should reflect the nuclear saturation properties.
In Figs. \ref{pot_o}(a) and \ref{pot_pb}(a), the variations of the ratios of the effective masses to their vacuum masses for the different mesons within the nucleus are presented. As the density increases towards the center, we observe the ratios are decreasing in a pattern where they are smaller for heavier mesons. These in-medium mass drops can be regarded as a measure of the role of light quarks in each meson or could be a reflection of partial chiral symmetry restoration.
This behavior resembles the nuclear matter properties and suggests that at very small radii, the effective masses obtained are consistent with the in-medium masses reported in neutron-rich nuclear matter \cite{krein1}.
The different vectorial interactions of the charged as well as neutral mesons within the nucleus, as stated in Eqs. (\ref{C2})-(\ref{C4}), are illustrated in Figs. \ref{pot_o}(b-d) and \ref{pot_pb}(b-d).  
The effects of larger repulsive $\o$ meson coupling are indicated in panels (b)-(d) of Figs. \ref{pot_o} and \ref{pot_pb}. 
From the behaviors of these mesic-nucleus potentials, the probable bound states can be speculated easily. It is noteworthy that positively charged mesons are excluded from our study due to the significant repulsion caused by the positive electric charge of the nucleus, which prevents the formation of such bound states. Moreover, the study in Ref. \cite{laura1} stated that it is highly unlikely to find $D^+$ mesic nucleus with an atomic number $Z\geq6$.
\subsection{Mesic-nuclei bound states} \label{4D}
 In this subsection, we present the bound state energies of the pseudoscalar mesons ($P$) in ${\rm{^{16}O}}$, ${\rm{^{40}Ca}}$, ${\rm{^{90}Zr}}$, ${\rm{^{197}Au}}$, and ${\rm{^{208}Pb}}$ nucleus for the potentials calculated using the QMC model. By analyzing the mesic-nucleus potentials from the previous subsection, the existence of the bound states can be predicted efficiently. In this study, the Klein-Gordon equation (\ref{final}) is solved in the coordinate space using the method described by Brockmann \cite{brock}. It is important to note that, the Eq. (\ref{final}) is solved for real potentials, assuming zero widths for the mesons both in free space and inside the nucleus. A negative binding energy and a well-behaved eigenstate within the nucleus for the given potentials are the signatures of the bound states. Under such conditions, the calculated meson-nucleus bound-state energies for various nuclei are listed below in Tables \ref{K},\ref{D},\ref{B}.
\begin{table}[th]
\centering
\begin{tblr}{colspec={cccccccc}}
\hline
&& ${\rm{^{16}_{8}O}}$ & ${\rm{^{40}_{20}Ca}}$ & ${\rm{^{90}_{40}Zr}}$ & ${\rm{^{197}_{79}Au}}$ & ${\rm{^{208}_{82}Pb}}$\\
\hline \hline
$K^-(V_\o^q)$ & $1s$& -81.67& -105.17& -112.39& -112.28& -100.75&\\
& $1p$& -42.95& -76.38& -92.81& -100.58& -90.02&\\
\hline
$K^-(V_\o^q,\;\rm{no\;Coulomb})$ & $1s$& -74.98& -94.36& -93.29& -87.91& -75.81&\\
& $1p$& -36.66 & -66.16 & -74.40 & -77.14 & -65.79&\\
\hline
$K^-(\tilde{V}_\o^q)$ & $1s$& -125.34 & -155.29& -158.45& -153.75& -136.48 &\\
& $1p$& -80.67 & -121.28& -136.56& -140.46& -123.96&\\
\hline
$K^-(\tilde{V}_\o^q,\;\rm{no\;Coulomb})$& $1s$& -118.54 & -144.29 & -139.18 & -129.12 & -111.32 &\\
& $1p$& -74.17 & -110.80& -117.89& -116.68& -99.49&\\
\hline
\hline
$\bar{K}^0(V_\o^q)$ & $1s$& -77.44& -104.52& -106.16& -106.30& -104.72&\\
& $1p$& -38.48 & -73.37 & -88.31& -93.84& -93.62&\\
\hline
$\bar{K}^0(\tilde{V}_\o^q)$ & $1s$& -121.25& -155.34 & -151.82& -147.99& -140.59&\\
& $1p$& -76.43 & -119.42 & -133.62& -133.97& -127.84&\\
\hline\hline
$K^0(V_\o^q)$ & $1s$& -2.52& -5.93& -5.37& -7.29& -6.74&\\
& $1p$& $\times$& $\times$& $\times$& -2.80& -2.55&\\
\hline
$K^0(\tilde{V}_\o^q)$ & $1s$& $\times$& $\times$& $\times$& $\times$& $\times$&\\
& $1p$& $\times$ & $\times$& $\times$& $\times$& $\times$&\\
\hline
\end{tblr}
\caption{ \raggedright{Binding energies (in MeV) for $K$-mesic–nucleus bound states in ${\rm{^{16}O}}$, ${\rm{^{40}Ca}}$, ${\rm{^{90}Zr}}$, ${\rm{^{197}Au}}$, and ${\rm{^{208}Pb}}$. We show the results for different conditions of nuclear and Coulomb interactions within the nuclei. The symbol `$\times$' refers to unbound states.}}
\label{K}
\end{table} 
Our results are limited to ground states, as higher states require consideration of spin-orbit interactions, which are beyond the scope of the current study.
The results show the possible formation of the bound states for $K(\bar{K})$ (in Table \ref{K}), $D(\bar{D})$ (in Table \ref{D}) and $B(\bar{B})$ (in Table \ref{B}) mesons with the ${\rm{^{16}O}}$, ${\rm{^{40}Ca}}$, ${\rm{^{90}Zr}}$, ${\rm{^{197}Au}}$, and ${\rm{^{208}Pb}}$ nucleus under the various conditions: (i) with the usual vector-isoscalar potential $V_\o^q$, (ii) with phenomenologically driven vector-isoscalar potential $\tilde{V}_\o^q$, (iii) with Coulomb interaction, and (iv) without Coulomb interaction.
The Coulomb interaction provides more attractive potentials to the negatively charged mesons within the nucleus, contributing approximately 6-7 MeV, 9-11 MeV, 18-19 MeV, 23-25 MeV, and 24-26 MeV binding for the $1s$ state for $K^-$, $D^-$, and $B^-$ mesons with ${\rm{^{16}O}}$, ${\rm{^{40}Ca}}$, ${\rm{^{90}Zr}}$, ${\rm{^{197}Au}}$, and ${\rm{^{208}Pb}}$ nucleus, respectively.
\begin{table}[th]
\centering
\begin{tblr}{colspec={cccccccc}}
\hline
&& $^{16}_{8}O$ & $^{40}_{20}Ca$ & $^{90}_{40}Zr$ & $^{197}_{79}Au$ & $^{208}_{82}Pb$\\
\hline \hline
$D^-(V_\o^q)$ & $1s$& -17.89& -22.27& -29.96& -33.18& -33.29&\\
& $1p$& -10.72 & -18.45& -25.49& -31.13& -31.60&\\
\hline
$D^-(V_\o^q,\;\rm{no\;Coulomb})$ & $1s$& -11.52& -12.45& -11.20& -10.14& -9.37&\\
& $1p$& -4.85 & -9.10& -7.68& -8.91& -8.22&\\
\hline
$D^-(\tilde{V}_\o^q)$ & $1s$& -2.78& -4.51& -10.66& -12.77& -16.69&\\
& $1p$& -2.25 & -3.95& -10.32& -12.51& -16.45&\\
\hline
$D^-(\tilde{V}_\o^q,\;\rm{no\;Coulomb})$& $1s$& $\times$& $\times$& $\times$& $\times$& $\times$&\\
& $1p$& $\times$ & $\times$& $\times$& $\times$& $\times$&\\
\hline
\hline
$\bar{D}^0(V_\o^q)$ & $1s$& -12.65& -18.38& -21.96& -26.24& -36.35&\\
& $1p$& -5.24& -12.15& -19.20& -23.30& -34.18&\\
\hline
$\bar{D}^0(\tilde{V}_\o^q)$ & $1s$& $\times$& $\times$ & $\times$& $\times$& -3.21&\\
& $1p$& $\times$ & $\times$ & $\times$& $\times$& -2.23&\\
\hline\hline
$D^0(V_\o^q)$ & $1s$& -103.96& -115.02& -107.74& -94.83& -82.38&\\
& $1p$& -88.48& -101.90& -100.21& -90.71& -78.08&\\
\hline
$D^0(\tilde{V}_\o^q)$ & $1s$& -151.71& -169.77& -155.65& -137.33& -119.49&\\
& $1p$& -135.06 & -152.43& -147.35& -132.31& -113.89&\\
\hline
\end{tblr}
\caption{ \raggedright{Same as Table \ref{K}, for D-mesic states.}}
\label{D}
\end{table} 
\begin{table}[th]
\centering
\begin{tblr}{colspec={cccccccc}}
\hline
&& ${\rm{^{16}_{8}O}}$ & ${\rm{^{40}_{20}Ca}}$ & ${\rm{^{90}_{40}Zr}}$ & ${\rm{^{197}_{79}Au}}$ & ${\rm{^{208}_{82}Pb}}$\\
\hline \hline
$B^-(V_\o^q)$ & $1s$& -119.77& -134.88& -130.71& -121.83& -110.11&\\
& $1p$& -113.62 & -124.60& -127.52& -119.39& -107.22&\\
\hline
$B^-(V_\o^q,\;\rm{no\;Coulomb})$ & $1s$& -112.73& -123.16& -110.82& -96.31& -84.16&\\
& $1p$& -106.67 & -113.31& -101.35& -94.81& -81.65&\\
\hline
$B^-(\tilde{V}_\o^q)$ & $1s$& -168.84& -193.55& -179.95& -165.04& -148.22&\\
& $1p$& -162.63 & -179.01& -176.69& -162.06& -144.43&\\
\hline
$B^-(\tilde{V}_\o^q,\;\rm{no\;Coulomb})$& $1s$& -161.79& -181.74& -160.01& -139.34& -122.15&\\
& $1p$& -155.64 & -167.49& -156.81& -136.80& -118.68&\\
\hline
\hline
$\bar{B}^0(V_\o^q)$ & $1s$& -115.66& -135.22& -122.03& -117.32& -114.61&\\
& $1p$& -109.45& -125.92& -118.73& -114.22& -111.59&\\
\hline
$\bar{B}^0(\tilde{V}_\o^q)$ & $1s$& -164.77& -193.33 & -171.11& -160.74& -152.78&\\
& $1p$& -158.51 & -180.32 & -167.67& -157.27& -149.15&\\
\hline\hline
$B^0(V_\o^q)$ & $1s$& -14.80& -14.43& -10.98& -10.70& -9.84&\\
& $1p$& -11.79& -13.40& -9.44& -10.37& -9.52&\\
\hline
$B^0(\tilde{V}_\o^q)$ & $1s$& $\times$& $\times$& $\times$& $\times$& $\times$&\\
& $1p$& $\times$ & $\times$& $\times$& $\times$& $\times$&\\
\hline
\end{tblr}
\caption{ \raggedright{Same as table- \ref{K}, for B-mesic states.}}
\label{B}
\end{table} 
\begin{figure}[th]
\centerline{\includegraphics[width=\textwidth]{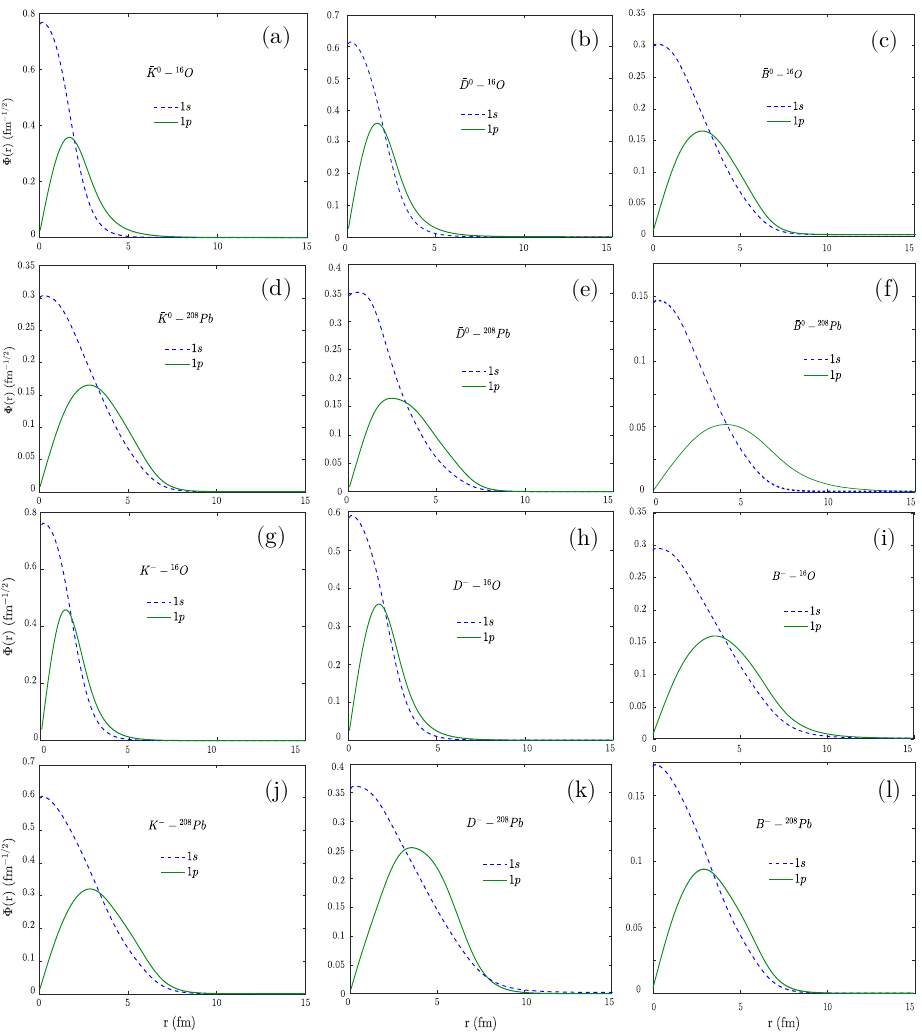}}
\vspace*{8pt}
    \caption{\raggedright{Bound state wave functions of the $\bar{P}^0$ ($\bar{K}^0$, $\bar{D}^0$, $\bar{B}^0$) and $P^-$ ($K^-$, $D^-$, $B^-$) mesons in ${\rm{^{16}O}}$ (a, b, c, g, h, i) and ${\rm{^{208}Pb}}$ (d, e, f, j, k, l) nuclei are presented. We show the results considering only normal vector coupling $V_\o^q$.}}
    \label{wf}
\end{figure}
Due to the sole contribution of the Coulomb force, the $D^-$ meson with the much more repulsive potential $\tilde{V}_\o^q$ can also form bound states.
This is due to the behavior of the wave functions for the potential $\tilde{V}_\o^q$, which indicates that the negatively charged $D^-$ mesons are comparatively pushed out of the nucleus and therefore the bound states are formed solely due to the Coulomb force.
With the same light quark structure, ${K}^0$ and ${B}^0$ mesons, since are blind to the Coulomb force, are unable to form the bound states for $\tilde{V}_\o^q$.
Therefore, the neutral states only experience the nuclear forces, and the resulting bound states are more immersed within the nucleus.
With the usual $\o$ coupling, the vector-isovector interaction produces a strong attractive potential for the $\bar{K}^0$ and $\bar{B}^0$ mesons compared to $\bar{D}^0$. 
The resulting binding energies become higher for the $\bar{K}^0$ and $\bar{B}^0$ mesons for a more repulsive $\o$ coupling ($\tilde{V}_\o^q$).
However, for such scenario, $\bar{D}^0$ is less likely to bound within nuclei.
It shows approximately 1 MeV, 6 MeV, 10 MeV, 16 MeV, and 26 MeV bindings are induced for the $\bar{D}^0$ meson while captured inside ${\rm{^{16}O}}$, ${\rm{^{40}Ca}}$, ${\rm{^{90}Zr}}$, ${\rm{^{197}Au}}$, and ${\rm{^{208}Pb}}$ nucleus, respectively, while for $\bar{K}^0$ and $\bar{B}^0$ mesons the obtained binding energies are quite large.
The values for $\bar{K}^0$ and $\bar{B}^0$ can be seen from Tables \ref{K} and \ref{B}, respectively.
Notably, with the repulsive $\o$ interaction, a few scenarios, which fail to form bound states, tend to form very light bound states with the heavy nuclei (e.g., ${\rm{^{208}Pb}}$).
In comparison to the results presented in Ref. \cite{D1}, the observed small binding energies are attributed to the inclusion of scalar-isovector potential. Essentially, the potential $V_\d$ induces opposite interactions between the mesons within the isospin doublets of $K(\bar{K})$, $D(\bar{D})$, and $B(\bar{B})$ mesons. As a consequence, it modifies the binding energies depending on the nature of their interactions. It is important to note that if the obtained states have small binding energies, they are more likely to mix with the continuum energy spectrum if their corresponding half-widths are greater \cite{laura1}. Therefore, this leads to indistinct states on the energy spectrum.
Furthermore, the $K^-$, $D^0$, and $B^-$ mesons form deeply bound states with the nuclei.
With $\tilde{V}_\o^q$, the bound-state energies are even more compared to those calculated with $V_\o^q$.
However, as indicated in Ref. \cite{D1}, due to the strong absorption of these states, the anticipated large width could make experimental observation difficult. 
The opposite vector-isoscalar ($\o$) interaction for $K^0\;(D^0, B^0)$ and $\bar{K}^0\;(\bar{D}^0,\bar{B}^0)$ leads to a significant difference in their binding energies, reflecting the presence of a strong Lorentz vector mean field. The results show the possible formation of bound states is more permissible with the heavier mesons, even when the potentials are comparable to each other. This represents the scalar interaction strengths of these mesons within the nuclei. 
Owing to the constituents of the open beauty mesons, they can form more strongly bound states with the nucleus. As depicted by the last column of Fig. \ref{wf}, for these states, the meson is much closer to the nuclei, which allows these states to probe even smaller variations in the nuclear matter properties \cite{krein1}.
\section{SUMMARY}
\label{summary}
To summarize we have investigated the probable formation of the mesic-nucleus bound states for various pseudoscalar mesons:$K\equiv(K^0, K^+)$, $\bar{K}\equiv(\bar{K}^0, K^-)$, $D\equiv(D^0, D^+)$, $\bar{D}\equiv(\bar{D}^0, D^-)$, $B\equiv(B^0, B^+)$ and $\bar{B}\equiv(\bar{B}^0, B^-)$ mesic-nucleus bound states in both light as well as heavy nuclei, specifically ${\rm{^{16}O}}$, ${\rm{^{40}Ca}}$, ${\rm{^{90}Zr}}$, ${\rm{^{197}Au}}$, and ${\rm{^{208}Pb}}$, within the quark meson coupling model. The model, extended with the realization of the Born-Oppenheimer approximation, allows one to study the properties of finite nuclei by accounting for the in-medium modifications of hadrons due to their internal structure responses to the nuclear environment. In this quark-based approach, scalar ($\sigma$, $\delta$) and vector ($\omega$, $\rho$) meson fields within the nucleus directly coupled to the light quarks and antiquarks of the mesons, resulting in modified meson properties. By solving the Klein-Gordon equation for real potentials, we have studied the formation of the bound states.
Our findings indicate the existence of mesic-nucleus bound states, where the meson-nucleus potentials provide substantial attraction to the corresponding mesons within the nucleus. The study suggests that heavier mesons are more likely to form strong bound states with heavier nuclei, particularly under (nearly) recoilless kinematics. The study of such bound states would significantly insights into strongly interacting systems, whereas, a precise measurement of the binding energies and width would provide valuable insight into their complex dynamics.
The upcoming experiments at ${\rm{\bar{P}ANDA}}$ at FAIR (GSI), JLab, and J-PARC are poised to investigate the nuclear medium properties through the formation of the bound states of mesons, where the current study of open strange and open heavy flavor mesons are expected to provide valuable insights for these investigations.

\end{document}